\documentclass{article}
\usepackage{arxiv}

\usepackage[utf8]{inputenc} 
\usepackage[T1]{fontenc}    
\usepackage{hyperref}       
\usepackage{url}            
\usepackage{booktabs}       
\usepackage{amsfonts}       
\usepackage{nicefrac}       
\usepackage{microtype}      
\usepackage{lipsum}         
\usepackage{graphicx}
\usepackage{natbib}
\usepackage{doi}
\usepackage{lineno}
\usepackage{amsmath}
\usepackage{comment}
\usepackage{graphicx}
\usepackage{xspace}
\usepackage[section]{placeins}
\usepackage{aas_macros}

\newcommand\at[2]{\left.\kern-\nulldelimiterspace#1\right|_{#2}}

\def\pc{\mathrm{pc}}
\def\cm{\mathrm{cm}} 

\def\km{\mathrm{km}} 
\def\kpc{\mathrm{kpc}}
\def\sec{\mathrm{s}} 

\let\oldequation\equation
\let\oldendequation\endequation
\renewenvironment{equation}
  {\linenomathNonumbers\oldequation}
  {\oldendequation\endlinenomath}

\let\oldalign\align
\let\oldendalign\endalign
\renewenvironment{align}
  {\linenomathNonumbers\oldalign}
    {\oldendalign\endlinenomath}

\title{One dimensional analytical solutions of the transport equations for incompressible magnetohydrodynamic (MHD) turbulence}
\date{}

\usepackage{authblk}

\newbox{\orcid}\sbox{\orcid}{\includegraphics[scale=0.06]{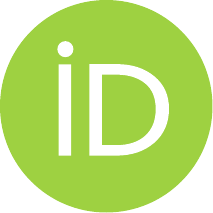}} 
\author[1]{%
	\href{https://orcid.org/0000-0002-6000-1262}{\usebox{\orcid}\hspace{1mm} Bingbing Wang \thanks{\texttt{bw0121@uah.edu}}}%
}
\author[1,2]{%
	\href{https://orcid.org/0000-0002-4642-6192}{\usebox{\orcid}\hspace{1mm} Gary P. Zank}%
}
\author[1,2]{%
	\href{https://orcid.org/0000-0003-1549-5256}{\usebox{\orcid}\hspace{1mm} Laxman Adhikari}%
}
\author[1]{%
	\href{https://orcid.org/0000-0001-8541-7523}{\usebox{\orcid}\hspace{1mm} Swati Sharma}%
}
\affil[1]{Center for Space Plasma and Aeronomic Research (CSPAR), University of Alabama in
Huntsville, Huntsville, AL 35899, USA}
\affil[2]{Department of Space Science, University of Alabama in Huntsville, Huntsville, AL 35899,USA}

\begin{document}

\maketitle
\begin{abstract}
We derive one dimensional (1D) analytical solutions for the transport equations of incompressible magnetohydrodynamic (MHD) turbulence developed by \citet{Zank2012, Adhikari2023}, including the Els\"{a}sser energies and the correlation lengths. 
The solutions are suitable for an arbitrary given background convection speed and Alfv\'en speed profiles but require near equipartition of turbulent kinetic energy and magnetic field energy. These analytical solutions provide a simple tool to investigate the evolution of turbulence and resulting energetic particle diffusion coefficients in various space and astrophysical environments that possess simple geometry.  
\end{abstract}


\section{Introduction}
Turbulence is important to a variety of astrophysical phenomena from diffusive shock acceleration to cosmic ray propagation, but there are few analytic solutions to turbulence transport models. 
The traditional Wentzel-Kramer-Brillouin (WKB) turbulence transport model describes the evolution of small amplitude incompressible fluctuations in a large-scale inhomogeneous flow \citep{Parker1965,Hollweg1973}. Although the analytical solution is accessible, the classic WKB model does not include the interaction between anti-propagating waves that initiates the dissipation of turbulence energy, nor can it describe the evolution of the correlation length. In an important paper \citet{Zank1996} clarified the relation between the WKB description of the fluctuating magnetic field variance and a simplified turbulence transport model that included the physics of dissipation and the corresponding correlation length. Unlike the WKB
model, \citet{Zank2012,Adhikari2023} derived much more complex transport equations based on the incompressible MHD equations that represent the leading-order description of  nearly incompressible MHD in the limit of large plasma beta. As the model consists of six coupled equations, it is far less tractable than the WKB model and is
only solved numerically \citep{Adhikari2016,Wang2022}.
In this work, by recognizing that the system of equations can be simplified to four equations by assuming near equipartition between the turbulent kinetic and magnetic field energy, we can derive 1D analytical solutions for the turbulence El\"asser energies and the corresponding correlation lengths. The solutions work for arbitrary convection speed and Alfv\'en speed profiles. A set of approximate solutions is also developed for turbulence with extremely weak dissipation or when the dissipation is balanced by some form of driving often manifested as plasma instabilities (examples already considered include turbulence generated by large-scale shear of different speed flows, shock waves, pick-up ion creation \citep{Zank1996,Zank2012,Zank2017}, cosmic ray streaming (\citep{Wang2022})).
These analytical solutions provide a simple and practical tool for studying the evolution of turbulence in various space and astrophysical environments, and are useful extensions of earlier analytical solutions with different simplifications by \citet{Zank1996,Adhikari2024b}.  

For the purposes of illustration, we apply the new analytic solutions of the turbulence transport equations to calculate the cosmic ray diffusion coefficient in the galactic halo. With the commonly adopted convection speed profile in cosmic ray studies \citep{Moskalenko2002,Porter2022}, we find a self-consistent cosmic ray diffusion coefficient, although the reality is likely to be much more complicated.
We derive the analytical solutions of the turbulence transport equations in Section 2, and as an application, apply the analytical solutions to investigate the cosmic ray diffusion coefficient in Section 3. Finally, we summarize our results in Section 4.

\section{1D analytical solution of the turbulence transport equation}
With the assumption that the turbulent kinetic energy nearly equals the turbulent magnetic field energy, the simplified 1D transport equations for leading order nearly compressible turbulence in a high plasma beta regime can be written as \citet{Zank2012,Adhikari2023} 
\begin{gather}
    (U-V_A)\frac{dz_+^2}{dx} + \left(\frac{1}{2}\nabla \cdot \mathbf U + \nabla \cdot \mathbf{V_A} \right) z_+^2 = -2\alpha\frac{z_+^2}{\lambda^+} (z_-^2)^{1/2} ; \\ 
    (U+V_A)\frac{dz_-^2}{dx} + \left(\frac{1}{2}\nabla \cdot \mathbf U - \nabla \cdot \mathbf{V_A} \right) z_-^2  = -2\alpha\frac{z_-^2}{\lambda^-} (z_+^2)^{1/2} ; \\ 
    (U-V_A)\frac{d\lambda^+}{dx} = 2\beta (z_-^2)^{1/2} ;\\
    (U+V_A)\frac{d\lambda^-}{dx} = 2\beta (z_+^2)^{1/2} ,
\end{gather}
where the fluctuating Els\"{a}sser variables $\mathbf{z}_{\pm}= \mathbf{u} \pm \mathbf{b}/\sqrt{\mu \rho}$, and $\mathbf u$, $ \mathbf{b}$, $\rho$ and $\mu$ are the turbulent velocity and magnetic fields (not assumed small but instead are the fluctuating components of a mean field decomposition), the mass density of the background plasma and the vacuum magnetic permeability, respectively, and $V_A$ is the Alfv\'en speed. $\lambda^{\pm}$ is the corresponding correlation length for $z_{\pm}$. $\alpha$ and $\beta$ are the von K\'arm\'an–Taylor constants. $z_{\pm}^2$ are Els\"asser energies. This set of equations is valid for 1D Cartesian coordinates or for a spherical coordinate system where $\mathbf U$ and $\mathbf{V_A}$ are along the radial direction.
By writing $z_{\pm}^2$ as the function of $\lambda^{\pm}$ through Equations (3,4) and substituting them into Equations (1,2), we obtain
\begin{gather}
    \frac{d^2\lambda^{\pm}}{dx^2} = -\left[ a_{\pm} + \frac{1}{\lambda^{\mp}} \frac{d\lambda^{\mp}}{dx} \right] \frac{d\lambda^{\pm}}{dx} ,\label{eq:dl2/dx2}
\end{gather}
after using the common assumption $\alpha = 2\beta$. Here $a_{\pm}$ has the form
\begin{equation}
a_{\pm} = \frac{\nabla(U \mp V_A)}{U\mp V_A} + \frac{\frac{1}{2}\nabla\cdot \mathbf{U} \mp \nabla \cdot \mathbf{V_A}}{2(U \pm V_A)}.
\end{equation}
Obviously, Equation (\ref{eq:dl2/dx2}) represents the full derivation of $\lambda^{\pm}\frac{d\lambda^{\mp}}{dx}$ and the solutions are given by
\begin{gather}
    \lambda^{+}\frac{d\lambda^{-}}{dx} = \lambda_0^{+}\at{\frac{d\lambda^{-}}{dx}}{x_0} \exp{\left( -\int_{x_0}^x a_{+}  dx \right)}; \label{eq:lpx} \\ 
    \lambda^{-}\frac{d\lambda^{+}}{dx} = \lambda_0^{-}\at{\frac{d\lambda^{+}}{dx}}{x_0} \exp{\left( -\int_{x_0}^x a_{-} dx \right)} , \label{eq:lmx}
\end{gather}
where the $\at{d\lambda^{\pm}/dx}{x_0}$ are given by Equations (3,4). 
Since $\lambda^+ \frac{d\lambda^{-}}{dx} + \lambda^- \frac{d\lambda^{+}}{dx} = \frac{d (\lambda^+ \lambda^-)}{dx}$, adding Equation (\ref{eq:lpx}) and (\ref{eq:lmx}) yields the expression for $\lambda^+\lambda^-$ as
\begin{equation}
    \lambda^+ \lambda^- = \lambda_0^+ \lambda_0^- + \lambda_0^{+}\at{\frac{d\lambda^{-}}{dx}}{x_0}  \int_{x_0}^x  \exp{\left(-\int_{x_0}^x a_{+} dx\right)}dx +  \lambda_0^{-}\at{\frac{d\lambda^{+}}{dx}}{x_0}  \int_{x_0}^x \exp{\left( -\int_{x_0}^x a_- dx\right)}dx . \label{eq:lpm} 
\end{equation}
By writing $\lambda^-=(\lambda^+ \lambda^-)/\lambda^+$, using Equation (\ref{eq:lpm}) and substituting it into Equation (\ref{eq:lmx}), we obtain 
\begin{gather}
    \frac{1}{\lambda^+} \frac{d\lambda^+}{dx} = \frac{\lambda_0^{-}\at{\frac{d\lambda^{+}}{dx}}{x_0} \exp{\left( -\int_{x_0}^x a_{-} dx \right)} }{\lambda^+ \lambda^-} ;   \label{eq:lp} \\
   \hspace*{-0.3cm} \lambda^+ = \lambda_0^+ \exp{\left( \int_0^x \frac{\lambda_0^{-}\at{\frac{d\lambda^{+}}{dx}}{x_0} \exp{\left( -\int_{x_0}^x a_{-} dx \right)} dx}{\lambda_0^+ \lambda_0^- +  \lambda_0^{+}\at{\frac{d\lambda^{-}}{dx}}{x_0}  \int_{x_0}^x  \exp{\left(-\int_{x_0}^x a_{+} dx\right)}dx +  \lambda_0^{-}\at{\frac{d\lambda^{+}}{dx}}{x_0}  \int_{x_0}^x  \exp{\left(-\int_{x_0}^x a_- dx \right)}dx }  \right)}. \label{eq:lpsol}
\end{gather}
Following the same method, we derive expressions for $\lambda^-$ as
\begin{gather}
    \frac{1}{\lambda^-} \frac{d\lambda^-}{dx} = \frac{\lambda_0^{+}\at{\frac{d\lambda^{-}}{dx}}{x_0} \exp{\left( -\int_{x_0}^x a_{+} dx \right)} }{\lambda^+ \lambda^-} ;   \label{eq:lp} \\
   \hspace*{-0.3cm} \lambda^- = \lambda_0^- \exp{\left( \int_0^x \frac{\lambda_0^{+}\at{\frac{d\lambda^{-}}{dx}}{x_0} \exp{\left( -\int_{x_0}^x a_{+} dx \right)} dx}{\lambda_0^+ \lambda_0^- +  \lambda_0^{+}\at{\frac{d\lambda^{-}}{dx}}{x_0}  \int_{x_0}^x  \exp{\left(-\int_{x_0}^x a_{+} dx\right)}dx +  \lambda_0^{-}\at{\frac{d\lambda^{+}}{dx}}{x_0}  \int_{x_0}^x  \exp{\left(-\int_{x_0}^x a_- dx \right)}dx }  \right)}. \label{eq:lpsol}
\end{gather}

The Els\"{a}sser energy $z_{\pm}^2$ can be derived from Equation (3,4) through
\begin{gather}
    z_{\pm}^2 = \frac{(U \pm V_A)^2}{4\beta^2} \left(\frac{d\lambda^{\mp}}{dx}\right)^2 . \label{eq:zpm}
\end{gather}
We have now obtained four exact analytical expressions for the Els\"{a}sser energy and associated correlation lengths.

For a special case in which the source of the turbulence is  balanced by the dissipation, the turbulence equations take the form $\alpha\to 0$. 
Equation (\ref{eq:lpsol}) then becomes
\begin{align}
    \begin{split}
        \lambda^+ & \approx \lambda_0^+ \exp\left( \int_{x_0}^x \frac{\lambda_0^- \at{\frac{d\lambda^+}{dx}}{x_0} \exp{\left(-\int_{x_0}^x a_- dx \right)}}{\lambda_0^+\lambda_0^-}\right) \\
        & \approx \lambda_0^+  +  \at{\frac{d\lambda^+}{dx}}{x_0} \int_{x_0}^x \exp{\left(\int_{x_0}^x -a_- dx \right)}dx ,
\end{split}
\end{align}
and $z_-^2$ is given by Equation (\ref{eq:zpm}) as
    \begin{equation}
        z_-^2 = \frac{(U-V_A)^2}{4\beta^2} \left(\frac{d\lambda^+}{dx}\right)^2 \approx z_{-0}^2 \left(\frac{U-V_A}{U_0-V_{A0}} \right)^2 \exp{\left(-2\int_{x_0}^x a_- dx\right)}. \label{eq:zp2}
\end{equation}
According to Equation (\ref{eq:lpx}), $d\lambda^-/dx$ can be written as 
\begin{align}
        \begin{split}
            \frac{d\lambda^-}{dx} & \approx \frac{ \lambda_0^+ \at{\frac{d\lambda^-}{dx}}{x_0} \exp{\left(-\int_{x_0}^x a_+ dx\right)}}{\lambda_0^++\at{\frac{d\lambda^+}{dx}}{x_0}\int_{x_0}^x \exp{\left(-\int_{x_0}^x a_-dx\right)}dx  } \\
            & \approx  \at{\frac{d\lambda^-}{dx}}{x_0} \exp{\left(-\int_{x_0}^x a_+ dx\right)}.
\end{split}
\end{align}
After integration over $x$, we obtain
\begin{gather}
    \lambda^- \approx \lambda_0^- + \at{\frac{d\lambda^-}{dx}}{x_0} \int_{x_0}^x \exp{\left(-\int_{x_0}^x a_+ dx\right)} dx.
\end{gather}
Like Equation (\ref{eq:zp2}), the solution for $z_+^2$ is
\begin{equation}
    z_+^2  \approx z_{+0}^2 \left(\frac{U+V_A}{U_0+V_{A0}} \right)^2 \exp{\left(-2\int_{x_0}^x a_+ dx\right)}.
\end{equation}

\section{A toy model for cosmic ray diffusion coefficient in the halo}
Based on the analytic solutions, we build a toy model to illustrate the nature of the cosmic ray diffusion coefficient in the galactic halo.
Assuming that the turbulent magnetic field in the interstellar medium follows a Kolmogorov power spectrum, the spatial diffusion coefficient ($D$) of cosmic ray particles is related to the strength of the turbulent magnetic field ($b^2$) and its correlation length ($\lambda_b$) as well as the large-scale magnetic field ($B$) through \citep{Zank1998,Wang2022}
\begin{equation}
    D \propto \left(\frac{B^{5/3}}{b^2} \right)  \lambda_b^{2/3} , \label{eq:D}
\end{equation}
where $b^2$ and $\lambda_b$ are given by \citep{Zank2017},
\begin{equation}
    b^2 = \mu\rho \frac{z_+^2+z_-^2}{4} ;\qquad \lambda_b = \frac{z_+^2\lambda^+ + z_-^2\lambda^-}{z_+^2+z_-^2}. \label{eq:deltaBl}
\end{equation}
Once the speed profiles for convection and Alfv\'en speed are determined, we can calculate the essential turbulence quantities and the resulting diffusion coefficient. Following cosmic ray propagation models \citep[see e.g.][]{Yuan2017}, the convection speed is assumed to increase linearly from the galactic disk to the halo, $U(x)= xdU/dx $, where $x$ is the distance from the galactic disk.
The Alfv\'en speed ($V_A = B/\sqrt{\mu \rho}$) is closely related to the momentum diffusion coefficient and the reacceleration process of low energy cosmic rays.
The mass density profile $\rho(x)$ can be obtained from the mass continuity equation 
\begin{equation}
    \rho(x) = \frac{\rho_0U_0}{U(x)} . \label{eq:rho}
\end{equation}
Thus, the Alfv\'en speed can be parametrized as $V_A = V_{A0}\sqrt{x/x_0}$.

We set the boundary at $x_0=0.1\,\kpc$ which is the height of the galaxy disk. For typical values $B=1.5$ $\mu G$ , $\rho_0/m_p=n_0=0.003\,\cm^{-3}$ ($m_p$ is the proton mass), $b_0^2/B^2=0.7$, we obtain $V_{A0} = 60\,\km\,\sec^{-1}$ and $z_{\pm}^2=5000\,(\km\,\sec^{-1})^2$. Assuming that the diffusion coefficient at $x_0$ is $4\times 10^{28}\cm^{2}/s$, the correlation lengths $\lambda^{\pm}_0$ are modeled as $40\,\pc$. 
The appropriate von Kármán–Taylor constant ($\alpha$) for the interstellar medium is unknown and can only be obtained by comparing the predicted and measured gas temperature, which is beyond the scope of this toy model. We consider a simple model where the damping of turbulence is balanced by the cosmic ray streaming instability \citep{Wiener2013,Wiener2017} or other possible coexisting instabilities; this formally corresponds to $\alpha \to 0$. Accordingly, in practice, we set $\alpha=1e-4$ in this work.

\begin{figure}[!hbt]
   \centering
   \includegraphics[width=.8\textwidth]{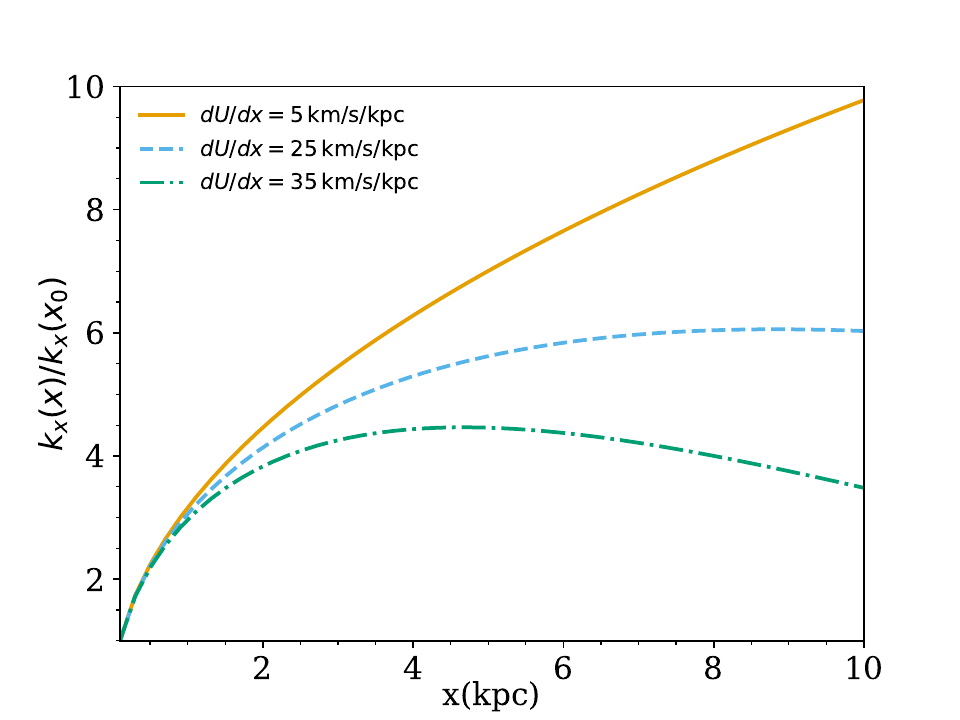} 
    \caption{Spatial dependence of the diffusion coefficient for three different gradients in flow speed. The diffusion coefficient is normalized by its value at the reference position $x_0=0.1\,\kpc$ as a function of distance to the galaxy disk. The flow speed is modeled as $U=x dU/dx$. 
    }
    \label{fg:1}
\end{figure}

According to Equation (\ref{eq:D}) and with Equation (\ref{eq:deltaBl}) and (\ref{eq:rho}), we can define the spatial dependence of the diffusion coefficient as
\begin{equation}
    k_x = x(z_+^2+z_-^2)^{-5/3}(z_+^2\lambda^++z_-^2\lambda^-)^{2/3}. 
\end{equation}

We illustrate the variation of $k_x$ with distance from the galactic disk ($x$) for three different choices of convection speed gradient ($dU/dx$) in Figure \ref{fg:1}. 
It can be seen that the diffusion coefficients near the galactic disk within $1\,\kpc$ increase rapidly at about the same rate. Thereafter, the diffusion coefficients continue to increase at slower rates until about $4\,\kpc$. The diffusion coefficient with the smaller
speed gradient increases faster than that with the larger gradient. Above $4\,\kpc$, the diffusion coefficient with a large speed gradient $dU/dx=35\,\km\,\sec^{-1}\,{\kpc}^{-1}$ (blue dash-dotted line) decreases with $x$. This contradicts the standard cosmic ray propagation model, where the diffusion coefficient is expected to be larger outside the cosmic ray diffusion halo to allow for the escape of cosmic rays.


\section{Summary} 
The six coupled transport equations for incompressible MHD turbulence can be simplified to four under the assumption of near equipartition of the turbulent kinetic and magnetic field energies. Exact 1D analytic solutions were derived and work for any specified convection and Alfv\'en speed. A set of approximate solutions for the case with ignorable or extremely weak turbulence dissipation was presented. A toy model is built to study the variation of the cosmic ray diffusion coefficient with distance from the galactic disk under the assumption that the turbulence dissipation is balanced by the cosmic ray streaming instability. 
Our aim is neither to fix the diffusion coefficient unambiguously nor to explore all possible parameter spaces. Rather, we consider a simple phenomenological scenario in order to illustrate a not completely unreasonable and self-consistent diffusion coefficient. This is crucial for developing a realistic spatially dependent cosmic ray propagation model. With this toy model, we found that the convection speed gradient should be sufficiently small to allow the diffusion coefficient to continue to increase with distance from the galactic disk.

\section*{Acknowledgments}
B.W, G.P.Z., and L.A. acknowledge the partial support of a NSF EPSCoR RII-Track-1 Cooperative Agreement OIA- 2148653, a NASA IMAP subaward under NASA contract 80GSFC19C0027, B.W, G.P.Z., L.A., and S.S. the partial support of a NASA Heliospheric DRIVE Center award SHIELD 80NSSC22M0164, and B.W. the partial support of a a NASA Heliospheric DRIVE Center SHIELD subaward 002134-00003.


\bibliographystyle{unsrtnat}
\bibliography{paper_analy.bib}

\end{document}